\documentstyle[12pt]{article}

\begin{document}

\title{BROWNIAN MOTION IN THE PRESENCE OF A TEMPERATURE GRADIENT}

\author{ {\bf A. P\'{e}rez-Madrid and J. M. Rub\'{\i}}\\Departament de
F\'{\i}sica Fonamental\\Facultat de F\'{\i}sica\\ Universitat de
Barcelona\\Diagonal 647, 08028 Barcelona, Spain\\ {\bf P. Mazur}\\
Instituut-Lorentz,\\ University of Leiden,\\ P. O. Box 9506, 2300 RA
Leiden, The Netherlands}

\date{}
\renewcommand{\theequation}{\arabic{section}.\arabic{equation}}

\maketitle
\parskip 2ex

\vspace{1.5cm}

\begin{abstract}

By considering an ensemble of Brownian particles suspended in a heat
bath as a thermodynamic system with an internal degree of freedom it
is possible to obtain the Fokker-Planck equation for Brownian motion
in a temperature gradient, by applying the scheme of non-equilibrium
thermodynamics. We recover explicitely the equations derived in
particular by Zubarev and Bashkirov using statistical mechanical and
kinetic methods. In addition when the temperature gradient does not
have an externally imposed magnitude we obtain the differential
equation for the temperature  field, which is coupled to the
Fokker-Planck equation.

\end{abstract}

\newpage

\renewcommand{\baselinestretch}{1.5}

\section{Introduction}

In 1965, Nicolis \cite{kn:Nicolis} considered the problem of Brownian motion in
a
temperature gradient by kinetic methods and derived an expression for
the thermal diffusion coefficient in this system. A few years later,
in 1968, Zubarev and Bashkirov \cite{kn:Zubarev} reconsidered this problem
using the methods developed by Zubarev for treating non-equilibrium systems.
They explicitely derive the Fokker-Planck equation for this case. In
addition they recovered Nicolis' expression for the thermal diffusion
coefficient. On the other hand Prigogine and Mazur
\cite{kn:Prigogine-Mazur}, \cite{kn:Mazur} introduced already
in 1953 the concept of internal degrees of freedom into the formalism
of non-equilibrium thermodynamics and pointed out that in this way
Fokker-Planck like equations could be obtained. Meixner
\cite{kn:Meixner} gave a more general framework for the derivation of
kinetic equations within this context.

In this paper we apply the non-equilibrium thermodynamic method of
internal degrees of freedom to Brownian motion in a temperature
gradient and thus derive in a simpler way the Fokker-Planck equation
for this case. We do not consider the temperature gradient as given
but establish also the coupled differential equation for the
temperature field. In section 2 we consider the Brownian particles
suspended in a nonuniform heat bath as a system with an internal
degree of freedom, the velocity $\vec{u}$ of the particles. We state
the corresponding Gibbs equation for the total differential of
entropy per unit volume and give the conservation laws for mass and
energy. The entropy balance equation is established in section 3,
where we also give expressions for the entropy flux and the entropy
source strength. In section 4 we follow the principles of non
equilibrium thermodynamics to formulate the linear phenomenological
laws for the dissipative fluxes occurring in the entropy source
strength, also for those pertaining to the internal coordinate space.
By substitution we then obtain the Fokker-Planck equation and the
coupled differential equation for the temperature field.
Finally, in section 5 we discuss the thermal-diffusion regime in which
the Fokker-Planck equation may be replaced by a simpler differential
equation for the density of the Brownian particles.

\section{The Brownian gas as a thermodynamic system with an internal
degree of freedom}
\setcounter{equation}{0}

Consider a dilute 'gas' of Brownian particles of mass $m$ suspended in a heat
bath at rest with constant mass density $\rho_H$. The Brownian gas
has an internal degree of freedom -internal coordinate- namely the
velocity $\vec{u}$ of a Brownian particle. The mass density of the
Brownian particles of velocity $\vec{u}$ at position $\vec{r}$ and
time $t$ in the heat bath is $\rho(\vec{u},\vec{r},t)$ while

\begin{equation}\label{eq:a1}
f(\vec{u},\vec{r},t) = \rho(\vec{u},\vec{r},t)/m
\end{equation}

\noindent is the probability density for the Brownian particles in
velocity-coordinate space, which may be viewed as an internal thermodynamic
variable.

The system, Brownian gas and heat bath, has mass density $\rho(\vec{r},t)$

\begin{equation}\label{eq:a2}
\rho = \rho_H + \rho_B = \rho_H + m \int f(\vec{u},\vec{r},t) d\vec{u},
\end{equation}

\noindent and energy $e(\vec{r},t)$ and entropy $s(\vec{r},t)$ per
unit of mass.

Since $\rho_H$ is constant, the Gibbs equation for this system can be
written as

\begin{equation}\label{eq:a3}
\delta \rho s = \frac{1}{T}\delta \rho e - \frac{m}{T}\int
\mu(\vec{u},\vec{r},t) \delta f(\vec{u},\vec{r},t) d\vec{u}
\end{equation}

\noindent where $\delta$ denotes the total differential of a quantity
and where $\mu(\vec{u},\vec{r},t)$ is the chemical potential per unit
mass of the Brownian gas component with internal coordinate
$\vec{u}$; $T(\vec{r})$ is the temperature of the heat bath at
position $\vec{r}$.

The chemical potential $\mu(\vec{u},\vec{r},t)$ is related to the
energy $e$ and entropy $s$ by

\begin{equation}\label{eq:a4}
\int \rho(\vec{u},\vec{r},t) \mu(\vec{u},\vec{r},t) d\vec{u} + \rho_H
\mu_H = \rho e - T\rho s + p
\end{equation}

\noindent Here $\mu_H$ is the chemical potential of the heat bath and
$p$ the system's hydrostatic pressure.

For the quantities $\rho(\vec{u},\vec{r},t) = f(\vec{u},\vec{r},t)/m$
and $e$, conservation laws may be written down. Let us first deal with
conservation of mass. For the rate of change with time of
$f(\vec{u},\vec{r},t)$, one may formally write

\begin{equation}\label{eq:a5}
\frac{\partial f}{\partial t} = - \dot{\vec{r}}\cdot \frac{\partial
f}{\partial \vec{r}} - \frac{\partial}{\partial \vec{u}}\cdot
\vec{J}_{\vec{u}} = -\vec{u}\cdot \frac{\partial f}{\partial \vec{r}}
- \frac{\partial}{\partial \vec{u}}\cdot \vec{J}_{\vec{u}}
\end{equation}

\noindent where $\vec{J}_{\vec{u}}$ represents a flux in velocity space which
results from the interaction with the heat bath (there is no external
field of force acting on the Brownian particles).

Integrating eq. (\ref{eq:a5}) over velocity $\vec{u}$ we obtain the
law of conservation of mass for the Brownian particles

\begin{equation}\label{eq:a6}
\frac{\partial \rho_B}{\partial t} = -\nabla \cdot \rho_B \vec{v}_B
\end{equation}

\noindent where $\rho_B$ is the mass density of the Brownian
particles, c.f. eq.(\ref{eq:a2}), and $\vec{v}_B(\vec{r},t)$,

\begin{equation}\label{eq:a7}
\vec{v}_B(\vec{r},t) = \rho_{B}^{-1} \int \rho(\vec{u},\vec{r},t)
\vec{u} d\vec{u},
\end{equation}

\noindent  their mean velocity.

The law of conservation of energy can be formulated as follows

\begin{equation}\label{eq:a8}
\frac{\partial \rho e}{\partial t} = -\nabla \cdot\vec{J}_q
\end{equation}

\noindent where $\vec{J}_q$ represent a heat flux defined in the
reference frame in which the heat bath is at rest.

In the next section we shall derive, using the Gibbs equation
(\ref{eq:a3}) and the conservation laws (\ref{eq:a5}) and
(\ref{eq:a8}), the entropy balance equation for the Brownian gas.

\section{Entropy balance and entropy source strength}
\setcounter{equation}{0}

Before deriving an expression for the entropy production in the
system described in section 2, we note that the Brownian gas, i. e.
the suspension of Brownian particles in the heat bath, may be
considered to be a many component ideal solution of 'components' with mass
density
$\rho(\vec{u},\vec{r},t)$ and that the thermodynamic potential
$\mu(\vec{u},\vec{r},t)$ is therefore of the form (with k Boltzmann's
constant)

\begin{equation}\label{eq:b1}
\mu(\vec{u},\vec{r},t) = \frac{kT}{m} ln f(\vec{u},\vec{r},t) +
C(\vec{u},\vec{r},t)
\end{equation}

\noindent The potential function $C(\vec{u},\vec{r},t)$ which depends
on the internal coordinate $\vec{u}$, can in principle still be a
function of the local thermodynamic state variables $T(\vec{r})$ and
$\rho_B(\vec{r},t)$.

Consider now the gas to be in local equilibrium, i.e. in the state of
internal equilibrium which would be reached if the system were
locally {\it insulated} at temperature $T(\vec{r})$ and mass density
$\rho_B(\vec{r},t)$. The chemical potential $\mu(\vec{u},\vec{r},t)$ and
the distribution function $f(\vec{u},\vec{r},t)$ must then satisfy
the following requirements: \\
\noindent 1. Since entropy at constant energy and constant density
of Brownian particles has in that state a maximum as a function of
the internal states $\rho(\vec{u},\vec{r},t)$, $\mu^{l.
eq.}(\vec{u},\vec{r},t)$ is uniform in velocity space and equal to
$\mu_B(\vec{r},t)$ the thermodynamic potential of a Brownian ideal
gas at temperature $T(\vec{r})$ and density $\rho_B(\vec{r},t)$

\begin{equation}\label{eq:b2}
\mu^{l. eq.}(\vec{u},\vec{r},t) = \mu_B \equiv \frac{kT}{m} \left(ln
\frac{\rho_B}{m} - \frac{3}{2} ln\frac{2\pi kT}{m} \right)
\end{equation}

\noindent 2. The distribution function $f(\vec{u})$ is Maxwellian at local
equilibrium and is given by

\begin{equation}\label{eq:b3}
f^{l. eq.}(\vec{u},\vec{r},t) = exp{ \left(\mu_B - \frac{1}{2}u^2\right)m/kT}
\end{equation}

\noindent If we then substitute eqs. (\ref{eq:b2}) and (\ref{eq:b3}),
for local equilibrium into eq. (\ref{eq:b1}), we find for $C(\vec{u})$

\begin{equation}\label{eq:b4}
C(\vec{u}) = \frac{1}{2}u^2
\end{equation}

\noindent and therefore that the chemical potential may be written as

\begin{equation}\label{eq:b5}
\mu(\vec{u},\vec{r},t) = \frac{kT}{m}ln f(\vec{u},\vec{r},t) +
\frac{1}{2}u^2,
\end{equation}

\noindent or equivalently

\begin{equation}\label{eq:b6}
\mu(\vec{u},\vec{r},t) = \mu_B + \frac{kT}{m} ln\left(f/f^{l. eq.}\right)
\end{equation}

Let us now calculate the rate of change of entropy per unit volume by
differentiating eq. (\ref{eq:a3}) with respect to time

\begin{equation}\label{eq:b7}
\frac{\partial \rho s}{\partial t} = \frac{1}{T}\frac{\partial \rho
e}{\partial t} - m\int \frac{\mu(\vec{u},\vec{r},t)}{T}\frac{\partial
f(\vec{u},\vec{r},t)}{\partial t}
\end{equation}

\noindent Using the conservation laws (\ref{eq:a5}) and (\ref{eq:a8}) as well
as eqs. (\ref{eq:b5}) and (\ref{eq:b6}), expressing $\mu(\vec{u},\vec{r},t)$ in
terms of $\rho(\vec{u},\vec{r},t) = mf(\vec{u},\vec{r},t)$, this equation may
be
written
in the form of a balance equation

\begin{equation}\label{eq:b8}
\frac{\partial \rho s}{\partial t} = -\nabla \vec{J}_s + \sigma
\;,\;\; \sigma>0
\end{equation}

\noindent where the entropy flux $\vec{J}_s$ and the entropy source
strength $\sigma$, which is a positive quantity in accordance with
the second law of thermodynamics, are given by

\begin{equation}\label{eq:b9}
\vec{J}_s = \vec{J}'_q/T - k\int f(\vec{u},\vec{r},t) (ln f(\vec{u},\vec{r},t)
-1)
\vec{u} d\vec{u}
\end{equation}

\begin{equation}\label{eq:b10}
\sigma = -\vec{J}'_q T^{-2} \nabla T - k \int \vec{J}_{\vec{u}}\cdot
\frac{\partial}{\partial \vec{u}} ln\left(f/f^{l. eq.}\right) d\vec{u}
\end{equation}

\noindent The modified heat flux $\vec{J}'_q$

\begin{equation}\label{eq:b11}
\vec{J}'_q = \vec{J}_q - m\int \frac{1}{2} u^2 \vec{u} f(\vec{u},\vec{r},t)
d\vec{u}
\end{equation}

\noindent contains a contribution to transfer of heat which is due to
the motion of the Brownian particles. In deriving eq. (\ref{eq:b9})
for the entropy flux use has been made of the identity

\begin{displaymath}
k \int \vec{u}\cdot \frac{\partial f}{\partial \vec{r}} ln f
d\vec{u} = k \frac{\partial}{\partial \vec{r}}\cdot \int \vec{u} flnf
d\vec{u} - k \int \vec{u}\cdot \frac{\partial
f}{\partial \vec{r}} d\vec{u}
\end{displaymath}

\begin{equation}\label{eq:b12}
= k\frac{\partial}{\partial \vec{r}}
\cdot \int \vec{u} f(ln f -1) d\vec{u}
\end{equation}

\noindent Furthermore, to obtain the second term in eq. (\ref{eq:b10}), a
partial
integration over velocity space has been performed, using the fact
that the flux $\vec{J}_{\vec{u}}$ vanishes as $\vec{u}\longrightarrow
\pm \infty$.

The entropy source strength $\sigma$ consists of two contributions:
the first represents the entropy created by heat conduction in the
heat bath in the presence of Brownian particles, the second, caused
by the motion of the Brownian particles in the heat bath, arises so to
say from diffusion in velocity space, the space of the internal
degree of freedom.

\section{Phenomenological relations and the Fokker-Planck equation
for Brownian motion in a temperature gradient}
\setcounter{equation}{0}

Following the principles of nonequilibrium thermodynamics \cite{kn:Mazur}, the
'linear' phenomenological relations -linear relations between the
fluxes and thermodynamic forces occuring in the entropy production
(\ref{eq:b10}) - are, since the system is isotropic and assuming
locality \cite{kn:Prigogine-Mazur}, \cite{kn:Mazur} in velocity space,

\begin{equation}\label{eq:c1}
\vec{J}'_q = -L_{TT} \nabla T/T^2 - \int k
L_{T\vec{u}}\frac{\partial}{\partial \vec{u}} ln\left(f/f^{l. eq.}\right)
d\vec{u}
\end{equation}

\begin{equation}\label{eq:c2}
\vec{J}_{\vec{u}} = -L_{\vec{u}T} \nabla T/T^2 - k
L_{\vec{u}\vec{u}}\frac{\partial}{\partial \vec{u}} ln\left(f/f^{l. eq.}\right)
\end{equation}

\noindent where the phenomenological coefficients obey the
Onsager-Casimir symmetry relations

\begin{equation}\label{eq:c3}
L_{\vec{u}T} = -L_{T\vec{u}}
\end{equation}

\noindent The assumption of locality in velocity space characterizes
the specific physical nature of the system considered and is
appropriate for Brownian motion.

With $\lambda = L_{TT}/T^2$ a heat conductivity coefficient and
defining the friction coefficients $\gamma$ and $\beta$

\begin{equation}\label{eq:c4}
\gamma \equiv L_{\vec{u}T}/fT \;\;, \;\;\;\;\beta\equiv mL_{\vec{u}\vec{u}}/fT
\end{equation}

\noindent relations (\ref{eq:c1}) and (\ref{eq:c2}), taking into
account relation (\ref{eq:c3}), become

\begin{equation}\label{eq:c5}
\vec{J}'_q = -\lambda \nabla T + m\int \gamma \left(f\vec{u} +
\frac{kT}{m} \frac{\partial f}{\partial \vec{u}}\right) d\vec{u}
\end{equation}

\begin{equation}\label{eq:c6}
\vec{J}_{\vec{u}} = -\gamma f \nabla T/T -  \beta \left(f\vec{u} +
\frac{kT}{m} \frac{\partial f}{\partial \vec{u}}\right)
\end{equation}

Considering $\gamma$ and $\beta$ to be independent of $\vec{u}$ in
first approximation and substituting eq. (\ref{eq:c6}) into the
conservation law (\ref{eq:a5}), we obtain the Fokker-Planck equation
for the Brownian motion in a heat bath with a non-uniform temperature
distribution

\begin{equation}\label{eq:c7}
\frac{\partial f}{\partial t} = -\vec{u}\cdot \frac{\partial
f}{\partial \vec{r}} + \beta \frac{\partial}{\partial \vec{u}}\cdot
\left(f\vec{u} + \frac{kT}{m}\frac{\partial f}{\partial
\vec{u}}\right) + \frac{\gamma}{T}\frac{\partial}{\partial
\vec{u}}\cdot f\frac{\partial T}{\partial \vec{r}}\;\;\;\;\cdot
\end{equation}

\noindent This equation is coupled to the differential equation
obtained by substituting eq. (\ref{eq:c5}) into the energy
conservation law (\ref{eq:a8})

\begin{equation}\label{eq:c8}
\frac{\partial \rho e}{\partial t} = \lambda \frac{\partial^2
T}{\partial r^2} - m\gamma \frac{\partial}{\partial \vec{r}}\cdot  \int
\left(f\vec{u} + \frac{kT}{m}\frac{\partial f}{\partial
\vec{u}}\right) d\vec{u} = \lambda \frac{\partial^2
T}{\partial r^2} - \gamma \frac{\partial}{\partial \vec{r}}\cdot
\rho_B \vec{v}_B
\end{equation}

\noindent In writing down eq. (\ref{eq:c8}) we have neglected the
small contribution to the heat flux arising from the kinetic energy
of the Brownian particles. Equation (\ref{eq:c7}) coincides, using their
definitions of constants and variables, with the Fokker-Planck
equation derived by Zubarev and Bashkirov by statistical mechanical
methods, except for one term. This term which these authors show to
be proportional to the effective volume of the Brownian particle and
as a consequence of negligible magnitude, can of course, due to its
extreme microscopic nature, not be found within the framework of a
thermodynamic theory. The two equations (\ref{eq:c7}) and
(\ref{eq:c8}) completely specify the coupled evolutions of the
temperature field and the velocity-coordinate probability
distribution of the Brownian particles. However after times much
larger than the characteristic time $\beta^{-1}$, the system enters the
diffusion and thermal diffusion regime for
which the evolution is governed by a simpler set of equations. We
shall discuss this regime in the next section.

\section{The thermal diffusion regime}
\setcounter{equation}{0}

Before discussing the equations which describe the long time behaviour of the
Brownian
particles and the heat bath, we shall derive the equation expressing, for the
Brownian
particles alone, the law of conservation of momentum, an equation which we did
not need
before. This equation will enable us to simplify the equation of motion  of
the Brownian gas for $t>>\beta^{-1}$ (see also in
connexion with the developments in this section ref.  \cite{kn:Mazur} ch X,
$\S$
8).

Using the definition (\ref{eq:a7}) as well as the continuity equation
(\ref{eq:a5}) for the distribution function $f(\vec{u},\vec{r},t)$ one
obtains the following equation of motion for the mean velocity
$\vec{v}_B(\vec{r},t)$:

\begin{equation}\label{eq:d1}
\rho_B \frac{d \vec{v}_{B}}{d t} = - \nabla\cdot
\vec{\vec{P}}_B(\vec{r},t) + m\int \vec{J}_{\vec{u}} d\vec{u}
\end{equation}

\noindent Here the hydrodynamic time derivation $d/d t$ for the
Brownian gas is defined as

\begin{equation}\label{eq:d2}
\frac{d}{d t} \equiv \frac{\partial}{\partial t} + \vec{v}_B \cdot
\frac{\partial}{\partial \vec{r}}
\end{equation}

\noindent while its pressure tensor $\vec{\vec{P}}_B$ is given by

\begin{equation}\label{eq:d3}
\vec{\vec{P}}_B(\vec{r},t) = m\int f (\vec{u} - \vec{v}_B) (\vec{u} -
\vec{v}_B) d\vec{u}
\end{equation}

\noindent Substituting into eq. (\ref{eq:d1}) the phenomenological
equation (\ref{eq:c6}) (which gave rise to the Fokker-Planck equation
(\ref{eq:c7})) the equation of motion becomes

\begin{equation}\label{eq:d4}
 \frac{d \vec{v}_{B}}{d t} + \rho_B^{-1}\nabla\cdot
\vec{\vec{P}}_B(\vec{r},t) + \gamma \nabla T/T = -\beta \vec{v}_B.
\end{equation}

As is well-known, in the diffusion regime, i.e. for
times long compared to the characteristic time $\beta^{-1}$, the
Brownian gas, due to collisions with the molecules of the heat bath,
will reach a state of internal equilibrium (i. eq.). In this state
the distribution function is approximately a Maxwellian corresponding
to a density $\rho_{B}(\vec{r},t)$ and temperature $T(\vec{r})$ and
a {\it non-vanishing mean velocity} $\vec{v}_{B}(\vec{r},t)$. It is then
given by

\begin{equation}\label{eq:d5}
f(\vec{u},\vec{r},t) \simeq f^{i. eq.}(\vec{u},\vec{r},t) \equiv
exp\{m[\mu_B - \frac{1}{2}(\vec{u} - \vec{v}_B)^2]/kT\}
\end{equation}

\noindent The pressure tensor $\vec{\vec{P}}_B$, eq. (\ref{eq:d3})
then reduces to the gas pressure $p_B$

\begin{equation}\label{eq:d6}
\vec{\vec{P}}_B = p_B \vec{\vec{U}}\;\; ,\;\;\;p_B = \rho_B kT/m,
\end{equation}

\noindent with $\vec{\vec{U}}$ the unit tensor. At the same time, the
inertia term at the left hand side of eq. (\ref{eq:d4}) becomes
negligibly small so that this equation, using also eq. (\ref{eq:d6}),
can be written, with $\vec{J}_D \equiv \rho_B \vec{v}_B$, as

\begin{equation}\label{eq:d7}
\vec{J}_D = - D \nabla \rho_B - D_T \nabla T/T
\end{equation}

\noindent where the diffusion coefficient D and the thermal diffusion
coefficient $D_T$  are defined respectively as

\begin{equation}\label{eq:d8}
D = \frac{kT}{m\beta}
\end{equation}

\begin{equation}\label{eq:d9}
D_T = \rho_{B}D\left(1 + \frac{\gamma m}{kT}\right)
\end{equation}

\noindent The form (\ref{eq:d9}) for $D_T$ agrees with the one obtained by
Zubarev and Bashkirov, and earlier by Nicolis, by other methods.

Let us finally discuss the form the entropy production $\sigma$, eq.
(\ref{eq:b10}), takes in the diffusion and thermal diffusion regime.  We have
stated that in this regime $f(\vec{u},\vec{r},t)$ is approximately given by eq.
(\ref{eq:d5}) and that in eq.  (\ref{eq:d4}), or the equivalent eq.
(\ref{eq:d1}), the inertia term may be neglected, while the pressure tensor
becomes diagonal. The following relation therefore holds in the diffusion
regime

\begin{equation}\label{eq:d10}
m \int \vec{J}_{\vec{u}} \;\;d\vec{u} = \nabla p_B
\end{equation}

\noindent and one then shows with eqs. (\ref{eq:b3}) and (\ref{eq:d5}) that
$\sigma$ reduces to

\begin{equation}\label{eq:d11}
\sigma = - \vec{J}_q \cdot \nabla T/T^2 - \vec{J}_D \cdot \nabla
p_{B}/\rho T
\end{equation}

\noindent

Using also eq. (\ref{eq:d6}), eq. (\ref{eq:d11}) may
be written as

\begin{equation}\label{eq:d12}
\sigma = - \vec{\tilde{J}}_{q} \cdot \nabla T/T^2 - \vec{J}_D \cdot
\frac{(k/m)\nabla\rho_B}{\rho_B}
\end{equation}

\noindent where a new modified heat flux $\vec{\tilde{J}}_{q}$ has been defined
by

\begin{equation}\label{eq:d13}
\vec{\tilde{J}}_{q} = \vec{J}_q + p_B \rho_{B}^{-1}\vec{J}_D
\end{equation}

\noindent Equation (\ref{eq:d12}), as does eq. (\ref{eq:d11}),
enables one to identify coupled thermodynamic fluxes and forces in
ordinary space for which Onsager symmetry relations must hold. Let us
then, with our previous results (\ref{eq:c5}) and (\ref{eq:d7}) for
$\vec{J}_q$ and $\vec{J}_D$ write down the resulting phenomenological
law for $\vec{\tilde{J}}_{q}$. This law is

\begin{equation}\label{eq:d14}
\vec{\tilde{J}}_{q} = - {\tilde{\lambda}} \nabla T - D_{T}T \frac{k}{m}
\frac{\nabla \rho_B}{\rho_B}
\end{equation}

\noindent Here the heat conductivity coefficient $\tilde{\lambda}$ at
uniform density of the Brownian gas is given by

\begin{equation}\label{eq:d15}
\tilde{\lambda} = \lambda + \frac{k}{m} \frac{D_{T}^2}{D\rho_B}
\end{equation}

\noindent We note that the coefficients occuring in eqs.
(\ref{eq:d7}) and (\ref{eq:d14}) have as expected the correct Onsager
symmetry.

 \vspace{1 cm}
\noindent{\bf \large ACKNOWLEDGMENTS}

\noindent This work has been partially supported by the CICYT of the Spanish
Government
under Grant PB92-0859.

\end{document}